\documentclass{article}
\usepackage{amsfonts}

\title{Classical models of affinely-rigid bodies with "thickness" in degenerate dimension}

\author{Vasyl Kovalchuk and Ewa Eliza Ro\.{z}ko\\
Institute of Fundamental Technological Research,\\
Polish Academy of Sciences,\\
21, \'{S}wi\c{e}tokrzyska str., 00-049 Warsaw, Poland\\
e-mails: vkoval@ippt.gov.pl, erozko@ippt.gov.pl}

\begin{document}

\maketitle
\begin{abstract}
The special interest is devoted to such situations when the material space of our object with affine degrees of freedom has generally lower dimension than the one of the physical space. In other words when we have the $m$-dimensional affinely-rigid body moving in the $n$-dimensional physical space, $m<n$. We mainly concentrate on the physical situation $m=2$, $n=3$ when "thickness" of flat bodies performs one-dimensional oscillations orthogonal to the two-dimensional central plane of the body. For the isotropic case in two "flat" dimensions some special solutions, namely, the stationary ellipses, which are analogous to the ellipsoidal figures of equilibrium well known in astro- and geophysics, e.g., in the theory of the Earth's shape, are obtained.
\end{abstract}


\section{Usual $n$-dimensional affinely-rigid bodies}

Let us consider the classical system of material points (discrete or continuous)
which we call the body \cite{JJS_82,JJS_book,all_04,all_05}. Let $(M,V,\rightarrow)$ be an affine space, where $M$ is a physical space in which our body is placed and $V$ is a linear space of translations (free vectors) in $M$. We may also introduce the metric tensor $g\in V^{\ast}\otimes V^{\ast}$ which makes our affine space a Euclidean one, i.e., $(M,V,\rightarrow,g)$. 

Let us suppose that we have labelled every material point of such a body in some way. Then let $(N,U,\rightarrow)$ be an affine space, where $N$ is the material space of such labels and $U$ is the corresponding linear space of translations in $N$. Similarly we may also introduce the metric tensor $\eta\in U^{\ast}\otimes U^{\ast}$ which makes our affine space a Euclidean one, i.e., $(N,U,\rightarrow,\eta)$. 

The position of the $a$-th material point at the time instant $t$ will be denoted by $x(t,a)$ ($x\in M,\ a\in N$) and an affine mapping from the material space into the physical one is as follows: 
\begin{equation}\label{eq1}
x^{i}(t,a)=r^{i}(t)+\varphi^{i}{}_{A}(t)a^{A},
\end{equation}
where $\varphi(t)$ is a linear part of the affine mapping ($\varphi$ is non-singular for any time instant $t$), i.e., $\varphi(t)\in LI(U,V)$, where $LI(U,V)$ is a manifold of linear isomorphisms from the linear space $U$ into the linear space $V$, $r(t)$ is the radius-vector of the centre of mass of our body if in the material space the position of the centre of mass is $a^{A}=0$. If the system is continuous, the label $a$ becomes the Lagrangian radius-vector (material variables) and $x$ becomes the Eulerian radius-vector (physical variables). In a wide range of problems there is no need to distinguish between continuous and discrete cases. We will use then the general form of description.

Thus, at any fixed $t\in \mathbb{R}$ the configuration space $Q$ of our problem is given by the following expression:
\begin{equation}\label{eq2a}
Q={\rm AfI}(N,M)=Q_{\rm tr}\times Q_{\rm int}=M\times 
{\rm LI}(U,V), 
\end{equation}
where tr and int refer to the translational and internal (relative) motion respectively (because the affine motion consists of spatial translations, rotations and homogeneous deformations; the last two are treated as the internal motion).
For $N=M=\mathbb{R}^{n}$ it becomes identical with the group space of the $n$-dimensional affine group:
\begin{equation}\label{eq2}
Q={\rm GAf}(n,\mathbb{R})\simeq\mathbb{R}^{n}\times_{s}{\rm GL}(n,\mathbb{R}),
\end{equation}
i.e., with the homogeneous space of this group with trivial isotropy groups. 

Then the considered system becomes an affinely-rigid body \cite{JJS_82,JJS_book,JJS-VK_03,all_04,all_05}, i.e., during any admissible motion all affine relations between constituents of the body are invariant (the material straight lines remain straight lines, their parallelism is conserved, and all mutual ratios of segments placed on the same straight lines are constant). The conception of the affinely-rigid body is a generalisation of the usual metrically-rigid body, in which during any admissible motion all distances (metric relations) between constituents of the body are constant (see, for example,~\cite{Arn_78,Gold_50}). 

In the special case of continuous medium the configuration space becomes a proper affine group:
\begin{equation}\label{eq3}
Q={\rm GAf}^{+}(n,\mathbb{R})\simeq\mathbb{R}^{n}\times_{s}{\rm GL}^{+}(n,\mathbb{R}),
\end{equation}
i.e., matrices $\varphi$ have positive determinants. For discrete systems the total affine group (\ref{eq2}) is in principle admissible.

\section{Affine bodies with degenerate dimension}

The above-described case is very popular and has wide spectrum of physical applications, but nevertheless there are problems when it is not suitable. So, the special interest in this article is devoted to such situations when the material space of our object with affine degrees of freedom has generally lower dimension than the one of the physical space:
\begin{equation}\label{eq4}
\dim N=m<n=\dim M. 
\end{equation}
In other words we have the $m$-dimensional affinely-rigid body moving in the $n$-dimensional physical space. Let us call such an object the affinely-rigid body with degenerate dimension \cite{Roz_05,Roz_PhD}.

Obviously, in physical applications only the special cases $n=3$ and $m=1,2$ are of direct interest but certain statements concerning general dimensions are also useful and should be quoted. We know that the standard continuum theory as well as some fundamental theories (e.g., the quantum field theory and the theory of elementary particles) deal with such objects as membranes, strings, etc. So, with the help of the notion of our affinely-rigid body with degenerate dimension some classical and quantum toy models of such objects may be defined.

Now the configuration space consists of affine injections, i.e., monomorphisms from the material space $N$ to the physical one $M$:
\begin{equation}\label{eq5}
Q=M\times{\rm LM}(U,V),
\end{equation}
where LM$(U,V)$ is the set of linear monomorphisms from $U$ to $V$. The formula (\ref{eq1}) is still valid but now the $n\times m$ matrix $\varphi^{i}{}_{A}$ has rank $m$.

Analytically, when we put $U=\mathbb{R}^{m}$ and $V=\mathbb{R}^{n}$, then
\begin{equation}\label{eq6}
Q=\mathbb{R}^{n}\times{\rm LM}(m,n).
\end{equation}
Our system has $f=n(m+1)$ degrees of freedom, i.e., $n$ translational and $nm$ internal (relative) ones. In this paper we are not particularly interested in the translational motion, therefore we investigate only the internal (relative) motion of our affinely-rigid body with degenerate dimension.

We see that $Q={\rm LM}(U,V)$ is a homogeneous space for the left-hand side action of GL$(V)$:
\begin{equation}\label{eq7}
A \in {\rm GL}(V):\qquad {\rm LM}(U,V) \ni \varphi \mapsto A\varphi \in {\rm LM}(U,V),
\end{equation}
but it fails to be homogeneous with respect to the material transformations, i.e., the right-hand side action of GL$(U)$:
\begin{equation}\label{eq8}
B \in {\rm GL}(U):\qquad {\rm LM}(U,V) \ni \varphi \mapsto \varphi B \in {\rm LM}(U,V),
\end{equation}
Indeed, this action is not transitive and its orbits consist of such $\varphi$'s which have the same images $\varphi(U)\in V$.

Thus, LM$(\mathrm{U},\mathrm{V})$ is a homogeneous space with respect to spatial transformations. It is interesting to describe this homogeneous space as a quotient manifold of GL$(\mathrm{V})$ with respect to some of its subgroup. Let us fix some standard linear monomorphism $\Psi$ of $\mathrm{U}$ into $\mathrm{V}$. So, we may say that if translational motion is neglected, $\mathrm{N}$ and $\mathrm{M}$ identified with $\mathrm{U}$ and $\mathrm{V}$ respectively, then LM$(\mathrm{U},\mathrm{V})$ may be obtained from $\Psi$ by the left actions:
\begin{equation}\label{eq9}
{\rm LM}(\mathrm{U},\mathrm{V}) \ni \Psi \mapsto \varphi = A \Psi
\in {\rm LM}(\mathrm{U},\mathrm{V}),
\end{equation}
where $A$ runs over GL$(\mathrm{V})$. The stabiliser subgroup $H [\Psi] \subset {\rm GL}(\mathrm{V})$ of the reference configuration $\Psi$ consists of those elements of GL$(\mathrm{V})$ which preserve not only the linear subspace $\Psi(\mathrm{U}) \subset \mathrm{V}$, but also any element of this subspace separately.

Let us describe these objects analytically. So, we put $U=\mathbb{R}^{m}$, $V=\mathbb{R}^{n}$, and identify $\mathbb{R}^{m}$ with linear subspace of $\mathbb{R}^{n}$ consisting of vectors with $m$ quite arbitrary entries on the first $m$ places and zeros on the remaining $(n-m)$ ones. Therefore
\begin{equation}\label{eq10}
\Psi\left(a^{1},\ldots,a^{m}\right)=
\left[a^{1},\ldots,a^{m},0,\ldots,0\right]^{T}=
\left[a^{1},\ldots,a^{m},\overline{o}\right]^{T},
\end{equation}
where $\overline{o}$ is the $(n-m)\times 1$ matrix built of zeros. It is easy to see that $H$ consists of all matrices of the following form:
\begin{equation}\label{eq11}
\left[
\begin{array}{cc}
\mathbb{I}_{m} & A \\
\mathbb{O} & B 
\end{array}\right],
\end{equation}
where $\mathbb{I}_{m}$ is the $m \times m$ identity matrix, $A$ and $B$ are
respectively $m\times(n-m)$ and $(n-m)\times(n-m)$ matrices, and $\mathbb{O}$ is the $(n-m)\times m$ matrix built of zeros. The matrices
$A$ and $B$ are arbitrary, or more precisely, subject only to the
restriction that the total matrix is non-singular (the set of
matrices violating this condition is obviously a measure-zero subset). The arbitrariness of $A$ and $B$ is respectively $m(n-m)$- and $(n-m)^{2}$-dimensional, therefore taken together they involve $n(n-m)$ arbitrary parameters. Thus, the quotient manifold GL$(V)/H$ is parameterized by $n^{2}-n(n-m)=nm$ essentially arbitrary parameters. In fact $\dim{\rm LM}(m,n)=\dim {\rm L}(m,n)=mn$. The fact that $H$ is indeed a subgroup of GL$(n,\bf
R)$ is a matter of direct calculation,
\begin{equation}\label{eq12}
\left[
\begin{array}{cc}
\mathbb{I} & A_{1} \\
\mathbb{O} & B_{1}
\end{array}\right]
\left[
\begin{array}{cc}
\mathbb{I} & A_{2} \\
\mathbb{O} & B_{2}
\end{array}\right]=
\left[
\begin{array}{cc}
\mathbb{I} & A_{1}B_{2}+A_{2} \\
\mathbb{O} & B_{1}B_{2}
\end{array}\right].
\end{equation}

For non-degenerate affine bodies (when $m=n$), one can define so-called affine velocity (Eringen's gyration) respectively in spatial and material representations, 
\begin{equation}\label{eq13}
\Omega=\dot{\varphi}\varphi^{-1}\in{\rm L}(V)\simeq{\rm GL}(V)^{\prime},\qquad \widehat{\Omega}=\varphi^{-1}\dot{\varphi}\in{\rm L}(U)\simeq{\rm GL}(U)^{\prime},
\end{equation}
and obviously $\Omega=\varphi\widehat{\Omega}\varphi^{-1}$. In degenerate models these objects do not exist. The right inverse of $\varphi$, i.e., $\varphi^{-1}_{\rm right}\in{\rm L}(V,U)$ satisfying 
\begin{equation}\label{eq14}
\varphi\varphi^{-1}_{\rm right}={\rm Id}_{V},
\end{equation}
does not exist at all because linear transformations cannot enlarge the dimension. Whereas the left inverses $\varphi^{-1}_{\rm left}\in {\rm L}(V,U)$, i.e., ones satisfying 
\begin{equation}\label{eq15}
\varphi^{-1}_{\rm left}\varphi={\rm Id}_{U},
\end{equation}
do exist but there is the infinite number of them (they coincide only on the subspace $\varphi(U)\in V$).

On the other hand the affine momentum (hypermomentum, affine spin) does exist. Its "laboratory" $\Sigma$ and "co-moving" $\widehat{\Sigma}$ representations are given as follows:
\begin{equation}\label{eq16}
\Sigma=\varphi p \in{\rm L}(V)^{\ast}\simeq{\rm
L}(V),\qquad \widehat{\Sigma}=p\varphi \in{\rm L}(U)^{\ast}
\simeq{\rm L}(U) 
\end{equation}
where $p \in{\rm L}(V,U)\simeq{\rm L}(U,V)^{\ast}$ is the canonical momentum conjugate to $\varphi$; the identification is made through the trace formula. Just as in the regular case, the components $\Sigma^{i}{}_{j}$ and $\widehat{\Sigma}^{A}{}_{B}$ are Hamiltonian generators of respectively the left action of GL$(V)$ and the right action of GL$(\mathrm{U})$ on LM$(U,V)$. Strictly speaking, the terms laboratory and co-moving representations are incorrect because there is nothing like the similarity transformation relating
$\Sigma$ to $\widehat{\Sigma}$. 

The doubled skew-symmetric part of $\Sigma$ and $\widehat{\Sigma}$ define respectively the spin and so-called vorticity:
\begin{equation}\label{eq17}
S^{i}{}_{j}:=\Sigma^{i}{}_{j}-g^{ik}g_{jl}\Sigma^{l}{}_{k},\qquad
\widehat{V}^{A}{}_{B}:=\widehat{\Sigma}^{A}{}_{B}-\eta^{AC}\eta_{BD}
\widehat\Sigma^{D}{}_{C}.
\end{equation}
They are Hamiltonian generators (momentum mapping) of the proper orthogonal subgroups SO$(V,g)\subset{\rm GL}(V)$ and SO$(U,\eta)\subset{\rm GL}(U)$ acting respectively on the left and right on LM$(U,V)$.

While $\varphi$ is not properly invertible, the contravariant tensor objects may
be transferred (pushed-forward) from the material space to the
physical one and the covariant tensors may be transferred from the physical space to the material one (pulled-back), but not conversely. 

\section{Flat bodies with non-zeroth "thickness"}

Later on we mainly concentrate on the two-dimensional body in the three-dimensional physical space. In contrary to the situation described in \cite{Roz_05}, where our body was infinitesimal in one dimension, in this article we study the model with non-zeroth "thickness". 

Generally speaking, kinematics and dynamics of such an object could be simply ones of the non-degenerate three-dimensional affinely-rigid body, but we would like to investigate the situation with the finite "thickness" when, in comparison to the non-degenerate case, additional constraints are implied, i.e., the "thickness" performs one-dimensional oscillations orthogonal to the two-dimensional central plane of the body. Then the group of material transformations has the following form:
\begin{equation}\label{eq18}
\mathbb{R}^{+} \times {\rm GL}\left(2,\mathbb{R}\right).
\end{equation}
The material space $\mathbb{R}^{3}$ is presented as Cartesian product $\mathbb{R}^{+} \times \mathbb{R}^{2}$, where $\mathbb{R}^{+}$ is the multiplicative group of positive numbers. The material transformations in $\mathbb{R}^{2}$ act as in \cite{Roz_05}, while $\mathbb{R}^{+}$ is the dilatation group in the third dimension.

We can identify configurations with the pairs $(\varrho,\varphi)$, where $\varphi$ describes the immersion of the central plane in the physical space, i.e., analytically $\varphi^{i}{}_{A}$ is the $3 \times 2$ matrix. An element $(k,B)$ acts on $(\varrho,\varphi)$ as follows:
\begin{equation}\label{eq19}
(k,B)\in\mathbb{R}^{+}\times{\rm GL}(2,\mathbb{R})\colon \qquad (\varrho, \varphi) \mapsto (k\varrho, \varphi B).
\end{equation}
We can represent the configuration directly with the help of the matrix
$\Phi:\mathbb{R}^{3} \rightarrow \mathbb{R}^{3}$. The conservation of orthogonality of the direction of dilatations to the central plane means that the matrix
\begin{equation}\label{eq20}
\Phi = \left[
\begin{array}{ccc}
  \Phi^{1}{}_{1} & \Phi^{1}{}_{2} & \Phi^{1}{}_{3} \\
  \Phi^{2}{}_{1} & \Phi^{2}{}_{2} & \Phi^{2}{}_{3} \\
  \Phi^{3}{}_{1} & \Phi^{3}{}_{2} & \Phi^{3}{}_{3} 
\end{array}
\right]
\end{equation}
fulfils the condition that third column has to be proportional to the vector product of first and second ones. If we consider $\Phi^{a}{}_{1}$, $\Phi^{b}{}_{2}$, $a,b=1,2,3$, as independent and arbitrary, then
\begin{equation}\label{eq21}
\Phi^{a}{}_{3}= \ell \ \varepsilon^{a}{}_{bc} \Phi^{b}{}_{1}
\Phi^{c}{}_{2},
\end{equation}
where $\varepsilon_{abc}$ is the completely antisymmetrical Ricci symbol, $\ell$ is the parameter which depends on the "thickness" and the variables describing the deformation in the central plane of the body, e.g., for the two-polar (singular value) decomposition (\ref{eq23}) we have that
\begin{equation}\label{eq22}
\ell_{\rm two-polar}=\frac{\varrho}{\lambda\mu},
\end{equation}
and the shifting of indices is understood in the trivial Kronecker-delta sense because we are working with the orthonormal coordinates. The above-described orthogonality is well known in the theory of plates and shells as the Kirchhoff-Love condition \cite{Love_96}.

{\bf Remark:} under this condition the affine group of the three-dimensional physical space does not act on the configuration space of the flat affinely-rigid body with "thickness". If we neglect the translational degrees of freedom, then the full linear group in three-dimensional translational space does not act on the configuration space of internal degrees of freedom. In fact,  these transformations in general violate above described orthogonality. On the other hand the action of Weyl group, i.e., the group of rotations combined with dilatations (in the sense of the physical space) in the configuration space is well defined. With respect to the action of this group our configuration space is not a homogeneous space because rotations and dilatations are not capable to deform the body in its central plane. On the other hand the configuration space still is the homogeneous space for the full group generated by $\mathbb{R}^{+}{\rm SO}(3,\mathbb{R})$, i.e., by rotations and dilatations in the physical space, and by GL$(2,\mathbb{R})$, i.e., by rotations and homogeneous deformations in the material space acting on the configurations from the right side.

\subsection{Two-polar decomposition}

In comparison to \cite{Roz_05} the two-polar (singular value) decomposition is written in the modified form, i.e., we use exclusively the $3 \times 3$ matrices:
\begin{equation}\label{eq23}
\Phi\left(\overline{k};\lambda,\mu,\varrho;\theta\right)= R\left(\overline{k}\right)D\left(\lambda,\mu,\varrho\right)U\left(\theta\right)^{-1},\qquad
\lambda,\mu,\varrho>0,
\end{equation}
where $R,U\in{\rm SO}(3,\mathbb{R})$ are proper orthogonal matrices and $D$ is diagonal, i.e.,
\begin{equation}\label{eq24}
D(\lambda,\mu,\varrho)=\left[
\begin{array}{ccc}
  \lambda & 0 & 0 \\
  0 & \mu & 0 \\
  0 & 0 & \varrho
\end{array}
\right],\quad U(\theta)^{-1}=\left[
\begin{array}{ccc}
  \cos\theta & \sin\theta & 0 \\
  -\sin\theta & \cos\theta & 0 \\
  0 & 0 & 1
\end{array}
\right].
\end{equation}

The co-moving angular velocities for $R$- and $U$-gyroscopes \cite{all_04,all_05} are as follows:
\begin{equation}\label{eq25}
\omega= R^{-1}\dot{R}=R^{T}\dot{R}=
\left[
\begin{array}{ccc}
0 & \omega _{3} & -\omega _{2} \\
-\omega _{3} & 0 & \omega _{1} \\
\omega _{2} & -\omega _{1} & 0
\end{array}
\right],\qquad {\omega}^{T}=-\omega,
\end{equation}
and
\begin{equation}\label{eq26}
\vartheta=U^{-1}\dot{U}=U^{T}\dot{U}=
\dot{\theta}\left[
\begin{array}{ccc}
  0 & -1 & 0 \\
  1 & 0 & 0 \\
  0 & 0 & 0 
\end{array}
\right], \qquad
\vartheta^{T}=-\vartheta.
\end{equation}
For $\dot{\Phi}$ and $\dot{\Phi}^{T}$ we have the following expressions:
\begin{equation}\label{eq27}
\dot{\Phi}=R\left(\dot{D}+\omega D-D\vartheta\right)U^{-1},\qquad
\dot{\Phi}^{T}=U\left(\dot{D}+\vartheta D-D\omega\right)R^{T}.
\end{equation}

The kinetic energy is assumed to have the usual form (we have only to substitute the constraints):
\begin{equation}\label{eq28}
T=\frac{1}{2}{\rm Tr}\left(J\dot{\Phi}^{T}\dot{\Phi}\right)=
\frac{1}{2}{\rm Tr}\left(U^{-1}JU\left[\dot{D}+\vartheta D-D\omega\right]\left[\dot{D}+\omega D-D\vartheta\right]\right),
\end{equation}
where $J\in U\otimes U$ is the twice contravariant, symmetric, non-singular, positively-definite tensor describing the inertial properties of our affinely-rigid body. If we take $J$ in the following diagonal form:
\begin{equation}\label{eq29}
J= \left[
\begin{array}{ccc}
  J_{1} & 0 & 0 \\
  0 & J_{2} & 0 \\
  0 & 0 & J_{3}
\end{array}
\right],
\end{equation}
then the above kinetic energy can be rewritten as follows:
\begin{eqnarray}
T&=&\frac{J_{1}\cos^{2}\theta+J_{2}\sin^{2}\theta}{2}\left(\frac{d\lambda}{dt}\right)^{2}+
\frac{J_{1}\sin^{2}\theta+J_{2}\cos^{2}\theta}{2}\left(\frac{d\mu}{dt}\right)^{2}
\nonumber \\
&+&\frac{J_{3}}{2}\left(\frac{d\varrho}{dt}\right)^{2}
+\frac{\left(J_{1}\sin^{2}\theta+J_{2}\cos^{2}\theta\right)\mu^{2}+J_{3}\varrho^{2}}{2}\ 
\omega_{1}^{2}\nonumber\\ 
&+&
\frac{\left(J_{1}\cos^{2}\theta+J_{2}\sin^{2}\theta\right)\lambda^{2}+J_{3}\varrho^{2}}{2}\ \omega_{2}^{2}
+\left(J_{1}+J_{2}\right)\lambda\mu\omega_{3}\frac{d\theta}{dt}\nonumber\\
&+&\left(J_{1}-J_{2}\right)\sin 2\theta
\left[\left(\mu\frac{d\mu}{dt}-\lambda\frac{d\lambda}{dt}\right)\frac{d\theta}{dt}+
\left(\lambda\frac{d\mu}{dt}-\mu\frac{d\lambda}{dt}\right)\omega_{3}+
\lambda\mu\omega_{1}\omega_{2}\right]\nonumber\\
&+&\frac{\left(J_{1}\cos^{2}\theta+J_{2}\sin^{2}\theta\right)\lambda^{2}
+\left(J_{1}\sin^{2}\theta+J_{2}\cos^{2}\theta\right)\mu^{2}}{2}\ \omega_{3}^{2}\nonumber\\
&+&\frac{\left(J_{1}\sin^{2}\theta+J_{2}\cos^{2}\theta\right)\lambda^{2}+
\left(J_{1}\cos^{2}\theta+J_{2}\sin^{2}\theta\right)\mu^{2}}{2}
\left(\frac{d\theta}{dt}\right)^{2}.\label{eq30}
\end{eqnarray}
Performing Legendre transformation we obtain that
\begin{eqnarray}
s_{1}=\frac{\partial T}{\partial \omega_{1}}&=&
\left[\left(J_{1}\sin^{2}\theta+J_{2}\cos^{2}\theta\right)\mu^{2}+
J_{3}\varrho^{2}\right]\omega_{1}+
\left(J_{1}-J_{2}\right)\sin 2\theta\lambda\mu\omega_{2},\nonumber\\
s_{2}=\frac{\partial T}{\partial \omega_{2}}&=&
\left[\left(J_{1}\cos^{2}\theta+J_{2}\sin^{2}\theta\right)\lambda^{2}+
J_{3}\varrho^{2}\right]\omega_{2}+
\left(J_{1}-J_{2}\right)\sin 2\theta\lambda\mu\omega_{1}, \nonumber\\ 
s_{3}=\frac{\partial T}{\partial \omega_{3}}&=&
\left[\left(J_{1}\cos^{2}\theta+J_{2}\sin^{2}\theta\right)\lambda^{2}
+\left(J_{1}\sin^{2}\theta+J_{2}\cos^{2}\theta\right)\mu^{2}\right]\omega_{3}\nonumber\\
&+&\left(J_{1}+ J_{2}\right)\lambda\mu\frac{d\theta}{dt}+
\left(J_{1}-J_{2}\right)\sin 2\theta\left[\lambda\frac{d\mu}{dt}-\mu\frac{d\lambda}{dt}\right],\nonumber\\
p_{\theta}=\frac{\partial T}{\partial \dot{\theta}}&=&
\left[\left(J_{1}\sin^{2}\theta+J_{2}\cos^{2}\theta\right)\lambda^{2}+
\left(J_{1}\cos^{2}\theta+J_{2}\sin^{2}\theta\right)\mu^{2}\right]\frac{d\theta}{dt}\nonumber\\
&+&\left(J_{1}+ J_{2}\right)\lambda\mu\omega_{3}+
\left(J_{1}-J_{2}\right)\sin 2\theta\left[\mu\frac{d\mu}{dt}-\lambda\frac{d\lambda}{dt}\right],\nonumber\\
p_{\lambda}=\frac{\partial T}{\partial \dot{\lambda}}&=&
\left(J_{1}\cos^{2}\theta+J_{2}\sin^{2}\theta\right)\frac{d\lambda}{dt}-
\left(J_{1}-J_{2}\right)\sin 2\theta\left[\lambda\frac{d\theta}{dt}+\mu\omega_{3}\right],\nonumber\\ 
p_{\mu}=\frac{\partial T}{\partial \dot{\mu}}&=&
\left(J_{1}\sin^{2}\theta+J_{2}\cos^{2}\theta\right)\frac{d\mu}{dt}+
\left(J_{1}-J_{2}\right)\sin 2\theta\left[\mu\frac{d\theta}{dt}+\lambda\omega_{3}\right],\nonumber\\ 
p_{\varrho}=\frac{\partial T}{\partial \dot{\varrho}}&=&
J_{3}\frac{d\varrho}{dt},\nonumber
\end{eqnarray}
where $s_{i}$ are canonical "spin" variables conjugate to angular velocities $\omega_{i}$.

\subsection{Isotropic case in two "flat" dimensions}

The above expressions simplify when we consider the isotropic case in two "flat" dimensions, i.e., when we have $J_{1}=J_{2}=J$. Then
\begin{eqnarray}
T&=& \frac{J}{2}\left[\left(\frac{d\lambda}{dt}\right)^{2}+
\left(\frac{d\mu}{dt}\right)^{2}\right]+
\frac{J_{3}}{2}\left(\frac{d\varrho}{dt}\right)^{2}+
\frac{J\mu^{2}+J_{3}\varrho^{2}}{2}\omega_{1}^{2}\nonumber\\
&+&\frac{J\lambda^{2}+J_{3}\varrho^{2}}{2}\omega_{2}^{2}+2J\lambda \mu \omega_{3}\frac{d\theta}{dt}+\frac{J}{2}\left(\lambda^{2}+\mu^{2}\right)
\left[\omega_{3}^{2}+\left(\frac{d\theta}{dt}\right)^{2}\right].\label{eq31}
\end{eqnarray}
Performing Legendre transformation we obtain that
\begin{eqnarray}
&&s_{1}=\left(J\mu^{2}+
J_{3}\varrho^{2}\right)\omega_{1},\qquad s_{2}=\left(J\lambda^{2}+J_{3}\varrho^{2}\right)\omega_{2},\nonumber\\
&&s_{3}=J\left(\lambda^{2}+\mu^{2}\right)\omega_{3}+
2J\lambda\mu\frac{d\theta}{dt},\qquad p_{\theta}=J\left(\lambda^{2}+\mu^{2}\right)\frac{d\theta}{dt}+
2J\lambda\mu\omega_{3},\nonumber\\
&&p_{\lambda}=J\frac{d\lambda}{dt},\qquad p_{\mu}=J\frac{d\mu}{dt},\qquad p_{\varrho}=J_{3}\frac{d\varrho}{dt}.\nonumber
\end{eqnarray}
After inverting this transformation, i.e.,
\begin{eqnarray}
&&\omega_{1}=\frac{s_{1}}{J\mu^{2}+ J_{3}\varrho^{2}},\qquad \omega_{2}=\frac{s_{2}}{J\lambda^{2}+ J_{3}\varrho^{2}},\nonumber\\
&&\omega_{3}=\frac{\left(\lambda^{2}+\mu^{2}\right)s_{3}-2\lambda\mu p_{\theta}}
{J\left(\lambda^{2}-\mu^{2}\right)^{2}},\qquad \frac{d\theta}{dt}=\frac{\left(\lambda^{2}+\mu^{2}\right)p_{\theta}-
2\lambda\mu s_{3}}{J\left(\lambda^{2}-\mu^{2}\right)^{2}},\nonumber\\
&&\frac{d\lambda}{dt}=\frac{p_{\lambda}}{J}, \qquad
\frac{d\mu}{dt} = \frac{p_{\mu}}{J}, \qquad \frac{d\varrho}{dt}=\frac{p_{\varrho}}{J_{3}},\nonumber
\end{eqnarray}
and substituting these expressions into (\ref{eq31}) we can rewrite the kinetic energy in the canonical variables:
\begin{eqnarray}
\mathcal{T}&=&\frac{s_{1}^{2}}{2\left(J\mu^{2}+J_{3}\varrho^{2}\right)}
+\frac{s_{2}^{2}}{2\left(J\lambda^{2}+J_{3}\varrho^{2}\right)}\nonumber\\
&+&\frac{\left(\lambda^{2}+\mu^{2}\right)\left(s_{3}^{2}+p_{\theta}^{2}\right)
-4\lambda \mu p_{\theta}s_{3}}{2J\left(\lambda^{2}-\mu^{2}\right)^{2}}
+\frac{p_{\lambda}^{2}+p_{\mu}^{2}}{2J}+\frac{p_{\varrho}^{2}}{2J_{3}}.
\label{eq32}
\end{eqnarray}

We can make an assumption that the potential $V$ depends only on the deformation, i.e., depends on $\Phi$ only through the Green deformation tensor
\begin{eqnarray}\label{eq33}
G&=&\Phi^{T}\Phi=UD^{2}U^{-1}\nonumber\\
&=&\left[
\begin{array}{ccc}
  \lambda^{2}\cos^{2}\theta+\mu^{2}\sin^{2}\theta & \left(\lambda^{2}-\mu^{2}\right)\sin\theta\cos\theta & 0 \\
  \left(\lambda^{2}-\mu^{2}\right)\sin\theta\cos\theta & \lambda^{2}\sin^{2}\theta+\mu^{2}\cos^{2}\theta & 0 \\
  0 & 0 & \varrho^{2}
\end{array}
\right],
\end{eqnarray}
which is not sensitive with respect to the left orthogonal mappings. We can introduce also the concept of deformation invariants $\mathcal{K}^{a}$, $a=\overline{1,3}$, which are scalar measures of deformation, basic stretchings, which do not contain any information concerning the orientation of deformation (its principal axes) in the physical or material space. They may be chosen in various ways but in an $n$-dimensional space exactly $n$ of them may be functionally independent. One of possible choices of the deformation invariants is to take the eigenvalues of the secular equation for the symmetric matrix $G$:
\begin{equation}\label{eq34}
\det\left[G-\mathcal{K}\mathbb{I}_{3}\right]=0,
\end{equation}
where $\mathbb{I}_{3}$ is the $3\times 3$ identity matrix, i.e.,
\begin{equation}\label{eq35}
\left[\mathcal{K}^{2}-
\mathcal{K}\left(\lambda^{2}+\mu^{2}\right)+\lambda^{2}\mu^{2}\right]
\left(\mathcal{K}-\varrho^{2}\right)=
\left(\mathcal{K}-\lambda^{2}\right)
\left(\mathcal{K}-\mu^{2}\right)\left(\mathcal{K}-\varrho^{2}\right)=0. 
\end{equation}
The deformation invariants are not sensitive with respect to both the spatial and material rigid rotations (isometries). So, the potential $V$ adapted to the two-polar decomposition is a function only of $\lambda$, $\mu$ and $\varrho$.

Moreover, one can see that in the kinetic energy expression (\ref{eq31}) the generalized velocities $\dot{\lambda}$, $\dot{\mu}$ corresponding to $\lambda$, $\mu$ and other variables describing the motion in the central plane of the body are separated from the generalized velocity $\dot{\varrho}$ describing the oscillations of the "thickness" $\varrho$. The same can be said also about the expression in the canonical variables (\ref{eq32}). The momentum $p_{\varrho}$ conjugated to $\varrho$ is orthogonal (in the sense of metrics encoded in the kinetic energy expression) to the other canonical momenta. Thus, the most simple are the dynamical models in which also the isotropic potential will have the separated form:
\begin{equation}\label{eq36}
V\left(\lambda,\mu,\varrho\right)=
V_{\lambda\mu}\left(\lambda,\mu\right)+V_{\varrho}\left(\varrho\right).
\end{equation}
As phenomenological models for $V_{\lambda\mu}$ we can use, for instance, the following ones:
\begin{eqnarray}
V^{1}_{\lambda\mu}\left(\lambda,\mu\right)&=&
\frac{k}{2}\left(\lambda^{2}+\mu^{2}\right),\qquad k>0,\label{eq37}\\
V^{2}_{\lambda\mu}\left(\lambda,\mu\right)&=&
c\left(\frac{1}{\lambda^2}+\lambda^2\right)+
d\left(\frac{1}{\mu^2}+\mu^2\right),\qquad c,d>0,\label{eq38}\\
V^{3}_{\lambda\mu}\left(\lambda,\mu\right)&=&
\kappa\left(\frac{1}{\lambda\mu}+\frac{\lambda^2+\mu^2}{2}\right),\qquad \kappa>0,\label{eq39}
\end{eqnarray}
and for the "thickness" potential $V_{\varrho}$ it can be, e.g.,
\begin{equation}\label{eq40}
V_{\varrho}(\varrho)= \frac{a}{\varrho}+ \frac{b}{2}\varrho^{2},\qquad a,b>0,
\end{equation}
which describes the nonlinear oscillations. The first term prevents from the unlimited compressing of the body, whereas the second one restricts the motion for large values of $\varrho$, i.e., prevents from the non-physical, unlimited stretching of the body.

So, the Hamiltonian (total energy) is written as 
\begin{equation}\label{eq41}
H=\mathcal{T}+V_{\lambda\mu}\left(\lambda,\mu\right)+V_{\varrho}(\varrho), 
\end{equation}
where $\mathcal{T}$ is taken in the form of (\ref{eq32}). 

The equations of motion can be calculated with the help of Poisson brackets as follows:
\begin{equation}\label{eq42}
\frac{ds_{i}}{dt}=\left\{s_{i},H\right\},\qquad
\frac{dp_{a}}{dt}=\left\{p_{a},H\right\},
\end{equation}
where $a=\left\{\theta,\lambda,\mu,\varrho\right\}$. The only non-zero basic Poisson brackets are
\begin{equation}\label{eq43}
\left\{q^{a},p_{b}\right\}=\delta^{a}{}_{b},\qquad 
\left\{s_{i},s_{j}\right\}=-\varepsilon_{ij}{}^{k}s_{k},
\end{equation}
and then we obtain that
\begin{eqnarray}
\frac{ds_{1}}{dt}&=&\frac{s_{2}}{J\left(\lambda^{2}-\mu^{2}\right)^{2}}
\left[\frac{J\mu^{2}\left(3\lambda^{2}-\mu^{2}\right)+
J_{3}\varrho^{2}\left(\lambda^{2}+\mu^{2}\right)}
{J\lambda^{2}+J_{3}\varrho^{2}}s_{3}-2\lambda\mu p_{\theta}\right],\
\label{eq44}\\
\frac{ds_{2}}{dt}&=&\frac{s_{1}}{J\left(\lambda^{2}-\mu^{2}\right)^{2}}
\left[\frac{J\lambda^{2}\left(\lambda^{2}-3\mu^{2}\right)-
J_{3}\varrho^{2}\left(\lambda^{2}+\mu^{2}\right)}
{J\mu^{2}+J_{3}\varrho^{2}}s_{3}+2\lambda\mu p_{\theta}\right],\label{eq45}\\
\frac{ds_{3}}{dt}&=&\frac{J\left(\mu^{2}-\lambda^{2}\right)s_{1}s_{2}}
{\left(J\lambda^{2}+J_{3}\varrho^{2}\right)\left(J\mu^{2}+J_{3}\rho^{2}\right)},
\qquad \frac{dp_{\theta }}{dt}=0,\label{eq46}\\
\frac{dp_{\lambda}}{dt}&=&-\frac{\partial V_{\lambda\mu}}{\partial\lambda}+
\frac{J\lambda s_{2}^{2}}{\left(J\lambda^{2}+J_{3}\varrho^{2}\right)^{2}}\nonumber\\
&+&
\frac{\lambda\left(\lambda^{2}+3\mu^{2}\right)\left(s_{3}^{2}+p^{2}_{\theta}\right)-
2\mu\left(\mu^{2}+3\lambda^{2}\right)s_{3}p_{\theta}}{J\left(\lambda^{2}-
\mu^{2}\right)^{3}},\label{eq47}\\
\frac{dp_{\mu}}{dt}&=&-\frac{\partial V_{\lambda\mu}}{\partial\mu}+\frac{J\mu s_{1}^{2}}{\left(J\mu^{2}+J_{3}\varrho^{2}\right)^{2}}\nonumber\\
&-&
\frac{\mu\left(\mu^{2}+3\lambda^{2}\right)\left(s_{3}^{2}+p^{2}_{\theta}\right)-
2\lambda\left(\lambda^{2}+3\mu^{2}\right)s_{3}p_{\theta}}{J\left(\lambda^{2}-
\mu^{2}\right)^{3}},\label{eq48}\\
\frac{dp_{\varrho}}{dt}&=&-\frac{dV_{\varrho}}{d\varrho}+\frac{J_{3}\varrho s_{1}^{2}}{\left(J\mu^{2}+J_{3}\varrho^{2}\right)^{2}}
+\frac{J_{3}\varrho s_{2}^{2}}{\left(J\lambda^{2}+J_{3}\varrho^{2}\right)^{2}}.\label{eq49}
\end{eqnarray}
The structure of the above expressions implies that even in the simplest case of the completely separated potential the dynamical coupling between the "thickness" parameter and the variables living in the central plane is present. 

\subsection{Stationary ellipses as special solutions}

Our equations of motion are strongly nonlinear and in general case there is hardly a hope to solve them analytically. Nevertheless, there exists a way to imaging some features of the phase portrait of such a dynamical system, i.e., we have to find some special solutions, namely, the stationary ellipses \cite{JJS_82,JJS_book}, which are analogous to the ellipsoidal figures of equilibrium well-known in astro- \cite{Bog_85} and geophysics, e.g., in the theory of the Earth's shape \cite{Chan_69}. 

In the case of the two-polar (singular value) decomposition (\ref{eq23}) we obtain then that the deformation invariants $\lambda$, $\mu$, $\varrho$ and the angular velocities $\omega$, $\vartheta$ are constant \cite{JJS_book}. If at the initial time $t=0$ we have that $\Phi_{0}=R_{0}\circ D\circ U^{-1}_{0}$, then at the time instant $t$ the configuration is as follows: 
\begin{equation}\label{eq50}
\Phi(t)=\left(R_{0}\circ e^{\omega t}\right)\circ D\circ \left(U_{0}\circ e^{\vartheta t}\right)^{-1}=R_{0}\circ e^{\omega t}De^{-\vartheta t}\circ U_{0}^{-1}.
\end{equation}
This can be rewritten in the following form:
\begin{equation}\label{eq55}
\Phi(t)=e^{\widehat{\omega} t}\circ \Phi_{0}\circ e^{-\widehat{\vartheta} t},
\end{equation}
where $\widehat{\omega}=R_{0}\circ \omega\circ R^{-1}_{0}$,
$\widehat{\vartheta}=U_{0}\circ \omega\circ U^{-1}_{0}$.

Therefore, the configuration $\Phi(t)$ we obtain from the initial configuration $\Phi_{0}$ acting on it by Euler, spatial isometry $e^{\widehat{\omega} t}\in {\rm SO}\left(V,g\right)$ and simultaneously acting on it by Lagrange, material isometry $e^{\widehat{\vartheta} t}\in {\rm SO}\left(U,\eta\right)$.

In our case from the special form of $\vartheta$ (\ref{eq26}) only one of the three branches described in \cite{JJS_book} is possible, i.e., when $\omega_{3}$, $d\theta/dt$ are constant, whereas $\omega_{1}=\omega_{2}=0$. Then the above equations of motions (\ref{eq44})--(\ref{eq49}) reduce as follows:
\begin{eqnarray}
\frac{\partial V_{\lambda\mu}}{\partial\lambda}&=&
\frac{\lambda\left(\lambda^{2}+3\mu^{2}\right)
\left(s_{3}^{2}+p^{2}_{\theta}\right)-
2\mu\left(\mu^{2}+3\lambda^{2}\right)s_{3}p_{\theta}}{
J\left(\lambda^{2}-\mu^{2}\right)^{3}},\label{eq52}\\
\frac{\partial V_{\lambda\mu}}{\partial\mu}&=&
\frac{2\lambda\left(\lambda^{2}+3\mu^{2}\right)s_{3}p_{\theta}-
\mu\left(\mu^{2}+3\lambda^{2}\right)\left(s_{3}^{2}+p^{2}_{\theta}\right)
}{J\left(\lambda^{2}-\mu^{2}\right)^{3}},\label{eq53}\\
\frac{dV_{\varrho}}{d\varrho}&=&0.\label{eq54}
\end{eqnarray}
We see that, while our parameters $s_{3}$, $p_{\theta}$ are completely arbitrary constant values, the above equations (\ref{eq52})--(\ref{eq54}) describe their interrelation with the deformation invariants $\lambda$, $\mu$, $\varrho$.  

{\bf Remark:} it is interesting to note that despite the stationary character of the solutions, the deformation is not constant in time. The Green deformation tensor (\ref{eq33}) as well as the Cauchy one, i.e., $C=\Phi^{-1T}\Phi^{-1}=RD^{2}R^{-1}$, depend on time explicitly through the time dependence of $U$ and $R$ respectively, i.e., 
\begin{equation}\label{eq51}
\frac{dG}{dt}=U\left(\vartheta D^{2}-D^{2}\vartheta\right)U^{-1}\neq 0,\quad \frac{dC}{dt}=R\left(\omega D^{2}-D^{2}\omega\right)R^{-1}\neq 0.
\end{equation}
Only the deformation invariants $\lambda$, $\mu$, $\varrho$ have the constant values, whereas the deformation tensors $G$ and $C$ perform the stationary rotations around their principal axis.

\section*{Summary}

The problem of dynamical systems on the manifolds of affine
injections is interesting in itself in the realm of analytical mechanics. 
But it is also applicable in the theory of structured bodies. Such bodies consist of small planes, like planar molecules and perhaps some supramolecular elements. Structured elements of degenerate dimension are known in condensed matter theory, let us mention, e.g., one-dimensional constituents of liquid crystals. Two-dimensional objects appear as three-atomic molecules like ${\rm H}_{2}{\rm O}$, ${\rm S}_{3}$, ${\rm CO}_{2}$ and molecules consisting of a larger number of molecules, but having some almost "flat" core. Also some applications in nanophysics are possible, but must be based on the quantum version of the theory. Some very interesting phenomena may appear there, because one deals in such problems with some convolution, overlap of the classical and quantum levels. Let us mention, e.g., flat or approximately flat molecules or the historical model of "Schwungrad" used in molecular dynamics in the pioneer days of quantum theory.

\section*{Acknowledgements}

This paper contains results obtained within the framework of the research project 501 018 32/1992 financed from the Scientific Research Support Fund in 2007--2010. Authors are greatly indebted to the Ministry of Science and Higher Education for this financial support.

One of authors (VK) is also very grateful to the Organizers of the Tenth International Conference on Geometry, Integrability and Quantization (June 6--11, 2008, Sts.\ Constantine and Elena, Varna, Bulgaria), especially to professor Iva\"{i}lo M.\ Mladenov, for their warm hospitality during the P-23/2006 "Group Structures Behind Two- and Three-Dimensional Elastic Structures" project visit within the framework of the agreement between Bulgarian and Polish Academies of Sciences (respectively, Institute of Biophysics and Institute of Fundamental Technological Research).

\end{document}